\documentclass[published]{JHEP3} 

\JHEP{00(2007)000}

\usepackage{epsfig,multicol,bbm}

\def\beq{\begin{equation}}
\def\eeq{\end{equation}}
\def\bea{\begin{eqnarray}}
\def\eea{\end{eqnarray}}

\title{A Theory of a Spot}

\author{Niayesh Afshordi\\
  Perimeter Institute for Theoretical Physics, Waterloo, Ontario N2L 2Y5, Canada, and\\
  Department of Physics and Astronomy, University of Waterloo, 200 University Avenue West, Waterloo, ON, N2L 3G1, Canada \\
  E-mail: \email{nafshordi@perimeterinstitute.ca}}

\author{An\v{z}e Slosar\\
  Brookhaven National Laboratory,
  Upton, NY 11973, USA\\
  E-mail: \email{anze@bnl.gov}}

\author{Yi Wang\\
   Physics Department, McGill University, Montreal, H3A 2T8, Canada \\
  E-mail: \email{wangyi@hep.physics.mcgill.ca}}

\received{\today} 		
\accepted{\today}		

\preprint{\hepth{9912999}}	

\abstract{We present a simple inflationary scenario that can produce
  arbitrarily large spherical underdense or overdense regions
  embedded in a standard $\Lambda$ cold dark matter paradigm, which we
  refer to as bubbles. We analyze the effect such bubbles would have
  on the Cosmic Microwave Background (CMB). For super-horizon sized bubble
  in the vicinity of the last scattering surface, a signal is imprinted
  onto CMB via a combination of Sach-Wolfe and an early integrated
  Sach-Wolfe (ISW) effects. Smaller, sub-horizon sized bubbles at
  lower redshifts (during matter domination and later) can imprint
  secondary anisotropies on the CMB via Rees-Sciama, late-time ISW and
  Ostriker-Vishniac effects. Our scenario, and arguably most similar
  inflationary models, produce bubbles which are over/underdense in
  potential: in density such bubbles are characterized by having a
  distinct wall with the interior staying at the cosmic mean density.
  We show that such models can potentially, with only moderate fine tuning,
  explain the \emph{cold spot}, a non-Gaussian feature identified in
  the Wilkinson Microwave Anisotropy Probe (WMAP) data by several
  authors. However, more detailed comparisons with current and future CMB data are necessary to confirm (or rule out) this scenario.}


\begin{document}

\section{Introduction}

Standard inflationary theory predicts that the primordial curvature
fluctuations are normally distributed
\cite{1981JETPL..33..532M,1982PhLB..115..295H,1982PhRvL..49.1110G,1982PhLB..117..175S,1983PhRvD..28..679B}. At
the level of linear perturbation theory this dictates that other
observables, such as temperature fluctuations in the cosmic microwave
background (CMB) should also be normally distributed. This has been
confirmed to hold in the real data with exquisite precision
\cite{2008arXiv0803.0732H,2008arXiv0803.0547K}.

However, a number of interesting features are present in the data that
seems to indicate a departure from the minimal inflationary scenario.
Wilkinson Map Anisotropy Probe (WMAP) \cite{2003ApJS..148....1B}
received a special attention from scientific treasure hunters, since
it is the only available full sky map of the CMB.  In addition to the
cold spot \cite{Cruz:2004ce} (discussed in more detail below), various
authors have found low multipole alignments
\cite{deOliveiraCosta:2006zj,Copi:2006tu,Land:2006bn,2003PhRvD..68f3501C},
north-south asymmetries \cite{2003astro.ph..7507E,Bernui:2005pz} and
various other non-Gaussian features found by blind searches
\cite{McEwen:2008kv,Wiaux:2007tc}. However, the statistical
significance of the results is difficult to ascertain, due to
publication bias and a-posteriori nature of some of the claims
presented in the literature
\cite{2004astro.ph..4567S,2004astro.ph..3073S}.

This paper is inspired by the so called \emph{cold spot}, a
nearly-circular region of the CMB that is allegedly too cold to be
generated by standard Gaussian inflationary perturbations. We discuss
a generic family of inflationary models/scenarios, as a generalized version of
multi-stream inflation \cite{Li:2009,Li:2009me}, which can produce large
spherical regions that have an offset in the potential with respect to
the surrounding universe. We enumerate possible effects that such
models would have on the cosmic microwave background.

These models are thus capable of reproducing the cold spot observed in
the CMB data, if the bubble is either in the line of sight to the
surface of last scattering or actually embedded in the surface of the
last scattering. We structure the paper as follows: In Section
\ref{sec:cold-spot-proposed} we discuss the detection of the cold spot
and the proposed explanations in the literature. In the subsequent
section \ref{sec:bubble-inflation} we develop our theory and show that
it can fit observations without excessive fine tuning (Section
\ref{sec:comp-with-observ} and Section \ref{sec:comp-cold-spot}).
Section \ref{sec:conclusions} concludes this article.

\section{Cold spot and proposed explanations}
\label{sec:cold-spot-proposed}

The cold spot was identified and named in \cite{Cruz:2004ce}, after
\cite{Vielva:2003et} found a detection of non-Gaussianity in the first
year WMAP data \cite{Bennett:2003bz} using Spherical Mexican Hat
Wavelets \cite{MartinezGonzalez:2001gu}. The detection significance
was about three sigma (about 0.1\% probability). The analysis was
performed on 15 arbitrary wavelet scales and two estimators, skewness
and kurtosis, were used. This indicates a dilution factor due to
a-posteriori detection by a factor of 30 (assuming tests are
uncorrelated), lowering the detection threshold to a bit over two
sigma. Later analysis confirmed this, and estimated that the a-priori
detection significance is about 1.9\%. \cite{Cruz:2006fy}. The cold
spot was also identified using other techniques: using more
sophisticated wavelet techniques \cite{Cayon:2005er}, steerable
wavelets \cite{Vielva:2007kt}, needlets \cite{Pietrobon:2008ve} and
scalar indices \cite{Rath:2007ti}. On the other hand
\cite{Zhang:2009qg} argue that the spot is not statistically
significant. Searches were performed looking for underdensity of
astrophysical objects at the location of the spot using radio sources
and galaxies and found no significant
detection\cite{Smith:2008tc,Granett:2009aw,Bremer:2010jn}.

The cold spot was found to be almost circular with angular radius of
about $4^{\circ}$, temperature decrement of $70 ~\mu K$, and
independent of the frequency \cite{Cruz:2006sv}.

Several different explanations have been proposed for the cold spot.
Perhaps the most natural is a large void between us and the last
scattering surface \cite{Inoue:2006rd,Inoue:2006fn}, which can create
the cold spot as a secondary anisotropy via the Rees-Sciamma effect.
Such void would have to be 200-300 Mpc/$h$ in size and have an
underdensity of $\delta\sim-0.3$ at redshift of $z\sim 1$. Such large
underdensities are extremely unlikely to appear spontaneously in the
standard scenario of structure formation.

Alternative proposals include cosmic texture
\cite{Cruz:2007pe,Cruz:2008sb}, chaotic post-inflationary preheating \cite{Bond:2009xx}, cosmic bubble collisions (e.g., \cite{Czech:2010rg}),  and nothing less than a gate to extra
dimensions \cite{Cembranos:2008kg}.

\section{Bubble blowing in multi-stream inflation}
\label{sec:bubble-inflation}
\begin{figure}
  \center
  \includegraphics[width=0.75\textwidth]{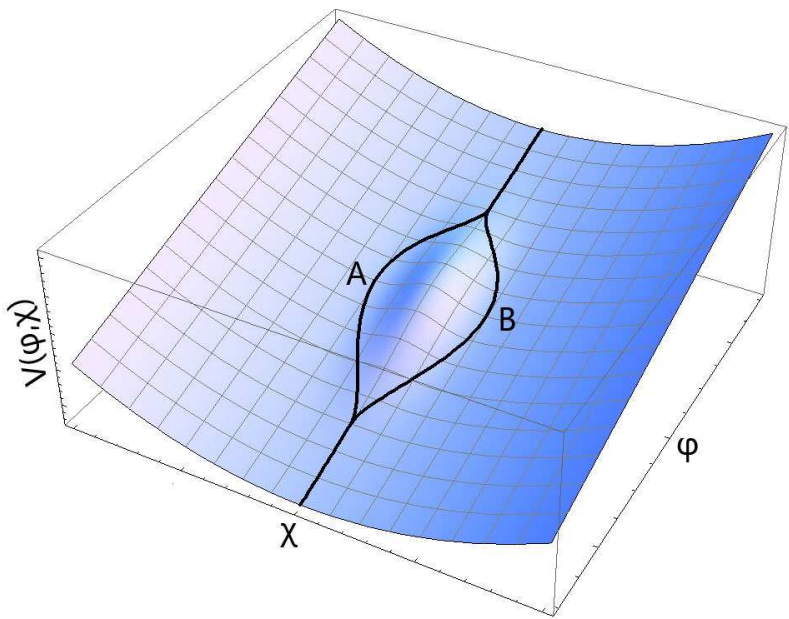}
    \caption{\label{fig:multi-stream}Multi-stream inflation.}
\end{figure}

It is clear that a mechanism that can create nearly circular bubbles or
overdensities can provide a plausible explanation of the cold
spot. To this end we consider a scenario in which the inflation can
experience different number of e-folds at different spatial
positions. This can be realized using several different scenarios. Our
first example is a two field inflation in which the minimum of the
inflaton potential causes spontaneous breaking of the mean-field
inflaton path into two distinct paths; an example of such potential is
illustrated in Figure \ref{fig:multi-stream}. A
similar result can be obtained if inflaton is able to quantum tunnel
e.g., from stream A to stream B in the top panel of Figure \ref{fig:multi-stream}.

\begin{figure}
  \center
  \includegraphics[width=0.85\textwidth]{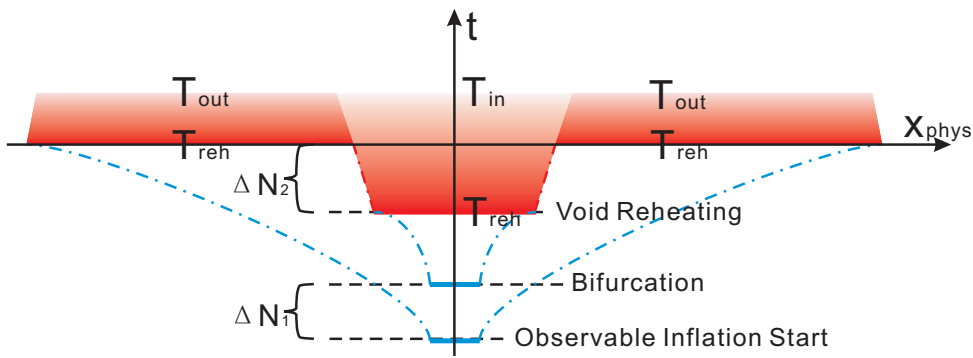}
    \caption{\label{fig:bubble} Expansion history for multi-stream inflation.}
\end{figure}

We can analyze such scenarios without constructing an explicit
inflationary model. The necessary physics is contained in three
parameters:
\begin{itemize}
\item Probability $p$ that the inflaton will bifurcate (or quantum
  tunnel) into the path $B$ rather than remaining on the main path
  $A$. For perfectly symmetrical bifurcation in the potential $p=0.5$,
  but note that this is not necessarily so.

\item $\Delta N_1$ is the number of e-foldings at which the bifurcation
  into paths $A$ and $B$ occurs, after the start of the
  observable inflation (i.e. $\sim 60$ e-foldings prior to the end of
  inflation).

\item $\Delta N_2$ is the total difference in the number of e-foldings
  between the two paths.
\end{itemize}
These are parameters illustrated in the Figure \ref{fig:bubble}.

In what follows, we will denote the regions that follows path $B$ as
bubbles, although the same mechanism can also produce overdense
spherical regions. 
To estimate the number density of such bubbles, we note, that at the
time of bifurcation, the number of causally connected regions in the
volume corresponding to the present-day observable universe is given
by $H_i^{-3}e^{3\Delta N_1}/H_b^{-3}$, where $H_i$ and $H_b$ are the
Hubble constants at the beginning of the observable inflation, and the
time of bifurcation. Therefore, the total number of bubbles  following that path $B$ of the inflation potential is
given by
\begin{equation}
  N_{\rm observable}=p ~(H_b/H_i)^3 e^{3\Delta N_1} = p~ e^{3(1-\epsilon) \Delta N_1} = p ~ e^{3\Delta N_1^*}~,\label{eq:n_obs}
\end{equation}
where we used the definition of the slow-roll parameter $\epsilon$:
\begin{equation}
\epsilon \equiv -\frac{d\ln H}{d\ln a}, \qquad \Delta N_1^* \equiv (1-\epsilon) \Delta N_1
\end{equation}
assuming that $\epsilon$ is roughly constant.

To calculate the physical size of the bubble, we note that the bubble
regions reheated $\Delta N_2$ e-foldings earlier and did not
exponentially expand
while reheating. The comoving radius of individual bubble regions is thus given by
\begin{equation}
   R_{\rm bubble}\sim \left(\frac{1}{H_0}e^{-\Delta N_1^*-\Delta N_2}\right).\label{eq:r_bubble}
\end{equation}

Finally, we want to calculate the density contrast between the bubble
and the rest of the universe. In this context, we treat the bubble
regions as perturbations in curvature and density. Since the curvature
perturbation is conserved outside horizon at all orders, our
calculation holds even for significantly underdense bubbles.

We start from the separate universe assumption. When we consider
super-Hubble perturbations, the universe within one Hubble volume is
locally an FRW universe, with metric
\begin{equation}
  ds^2={\cal N}({\bf x},t)^2 dt^2-a^2(t)e^{-2\psi({\bf x},t)}d{\bf x}\cdot d{\bf x},\label{metric}
\end{equation}
where ${\cal N}$ and $\psi$ quantify the local physical time/distance
with respect to the temporal and spatial coordinates
\cite{Afshordi:2000nr,Lyth:2004gb}.  If perturbations are adiabatic
(i.e. pressure is a function of density), energy conservations implies
that the curvature perturbation of the uniform density slices :
\begin{equation}
  \zeta({\bf x})=\psi({\bf x},t)-\frac{1}{3}\int_{\rho_b(t)}^{\rho({\bf x},t)}\frac{d\rho'}{\rho'+p(\rho')}~,\label{Bardeen}
\end{equation}
is conserved and gauge invariant, where $\rho_b(t)$ and $a(t)$ in
Eqs. (\ref{metric})-(\ref{Bardeen}) are the background (unperturbed)
values of density and scale factor.  Moreover, $p(\rho)=\rho/3$ after
reheating, and $p=0$ when matter dominates over radiation. One can
further show that $\zeta$ equals to the e-folding number difference
between an uniform density slice and a flat slice. The e-folding
number in the flat slice is the same as the background e-folding
number. Note that the reheating surface is a uniform density slice, so
on the reheating surface (the surface with temperature $T_{\rm reh}$
in Fig. \ref{fig:bubble}), we have
\begin{equation}
 \zeta_* = -(N_{\rm bubble}-N_0)=\Delta N_2~.
\end{equation}
Note that since $\zeta_*$ is a constant, the primordial bubble profile is a top-hat. Moreover, $\zeta_*$ is gauge invariant. So in a flat slice, we have
\begin{equation}
  \delta \rho=\left( e^{-3(1+w)\Delta N_2}-1 \right)\rho~,
\end{equation}
where $w=1/3$ when radiation dominates, and $w=0$ when matter
dominates. Also, notice that when the bubble reheats earlier, we have $\Delta
N_2 >0$. In this case, $\delta\rho<0$. In other words, we are
producing underdense regions.

Another important point is that for rare bifurcation events, $p \ll
1$, we expect nearly spherical bubbles. The reason is that the rare
peaks of a random gaussian field with a fixed threshold  tend to be
spherical (e.g., \cite{Bardeen:1985tr}). In our case, the random gaussian field could be the transverse fluctuations about the inflationary trajectory which need to be large to trigger bifurcation.

\section{Effects of cosmic bubbles on the CMB}
\label{sec:comp-with-observ}

For the calculation of CMB anisotropies, we shall use the linear
metric in the Newtonian (or longitudinal) gauge: \beq ds^2 =
a^2(\eta)\left\{[1+2\phi({\bf x},\eta)]d\eta^2 - [1-2\psi({\bf
    x},\eta)]d{\bf x}\cdot d{\bf x}\right\}, \eeq where $\phi({\bf
  x},\eta)$ and $\psi({\bf x},\eta)$ are expressed in terms of
comoving coordinates ${\bf x}$ and conformal time $\eta$, which is
related to proper time through $dt = a(\eta)d\eta$ (see
\cite{Mukhanov:1990me} for comprehensive review of cosmological
perturbation theory). Since $\phi = \psi + {\cal O}(\psi^2)$ for
perfect fluids, for most of what follows (with the notable exception
of Rees-Sciama effect), we use $\phi \simeq \psi$.

The gauge-invariant curvature perturbation: \beq \zeta = \psi -
\frac{H}{\dot{H}}(H\psi+\dot{\psi}) = \psi +
\frac{2(\psi+\dot{\psi}/H)}{3(1+w)},\label{zeta_l} \eeq is constant on
superhorizon scales during the radiation era, and on all scales in the
matter era, and can be derived from Eq. (\ref{Bardeen}) (for linear
perturbations) using the linearized $G_{00}$ Einstein equation on
superhorizon scales: \beq -3H(H\psi+\dot{\psi}) = 4\pi G
\delta\rho, \label{G00} \eeq and Friedmann equation \beq \dot{H} =
-4\pi G(\rho+p).  \eeq

Therefore, for the linear Fourier modes $\psi_{\bf k}$ relevant for
the cold spot we can write: \beq \psi_{\bf k}(t) = g(t) T(|{\bf k}|)
\zeta_{\bf k}, \label{psi_k} \eeq where $T(k)$ is the transfer
function for matter density which goes to $1$ on scales bigger than
the comoving horizon at matter-radiation transition and $g(t)$ is the
homogeneous solution to the equation of $\zeta$ conservation
(Eq. \ref{zeta_l}), assuming that $\dot{g}=0$ deep into the radiation
era: \beq g+\frac{2(g+\dot{g}/H)}{3(1+w)} =1.  \eeq During transition
between radiation and matter dominated universes, this equation can be
solved analytically giving
\beq g(a) = \frac{3}{5} + \frac{2}{15}u - \frac{8}{15}u^2
+\frac{16}{15}u^3 \left(\sqrt{1+u^{-1}}-1 \right), \label{eq:g_analytic}\eeq where
$u=\Omega_{\rm rad}/\Omega_m a=a_{\rm eq}/a$. Hence, $g(t)$ goes from
$2/3$ in radiaton era to $3/5$ in the matter era, and then decays when
dark energy starts to dominate. In the transition from matter
domination to dark energy domination, the equation must be solved numerically.

\subsection{Superhorizon bubbles: Sachs-Wolfe and early ISW effects}

To be detectable in the CMB, the bubble should have a superhorizon
size if it is close to the last scattering surface. This significantly
simplifies the calculations, as the curvature perturbation $\zeta$
remains constant. The CMB anisotropies are limited to Sachs-Wolfe and
early Integrated Sachs-Wolfe (ISW) effects \cite{Sachs:1967er}. The
latter is due to the fact that close to the matter-radiation
transition the Newtonian potential is varying. The anisotropies are
given by: \beq \frac{\delta T(\hat{\bf n})}{T} = \psi(r_{LSS}\hat{\bf
  n},\eta_{LSS})+\Theta (r_{LSS}\hat{\bf n},\eta_{LSS})+2
\int_{\eta_{LSS}}^{\eta_{0}} d\eta \frac{\partial\psi(\hat{\bf
    n}(\eta_0-\eta),\eta)}{\partial \eta},\label{SW+ISW} \eeq where
$\Theta({\bf x},\eta)$ is the intrinsic temperature fluctuation in the
photon field, while $\eta_0$ and $\eta_{LSS}$ are the conformal times
at present and last scattering respectively. Notice that partial
derivative in Eq. (\ref{SW+ISW}) is with respect to conformal time at
fixed position, while the integral is over the past light cone, which
is why it cannot be taken trivially.

\begin{figure}
  \centering
  \includegraphics[width=\linewidth]{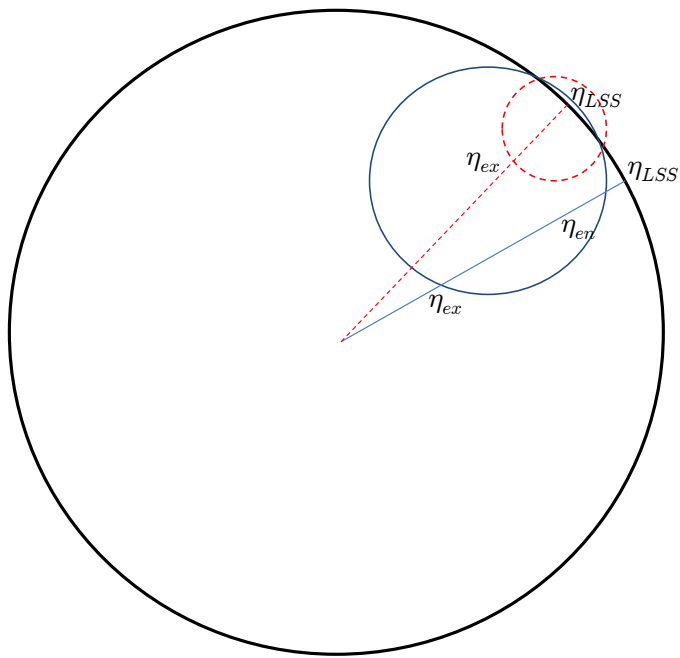}
\caption{Schematic picture of two bubbles that intersect the last scattering surface. Solid and dashed lines show two typical lines of sight that intersect these bubbles.$\eta_{LSS}$, $\eta_{en}$, and $\eta_{ex}$ are conformal times at last scattering, as well as the entry and exit intersections of the line of sight with the bubble. 
\label{fig:LSS}}

\end{figure}

In order to find $\Theta$, we can combine the adibaticity conditions:
\beq \delta\rho_m = 3\rho_m\Theta, \delta\rho_{rad} =
4\rho_{rad}\Theta, \eeq with the superhorizon $G_{00}$ equation
(\ref{G00}), which yields: \beq \Theta =
-\frac{2(\psi+\dot{\psi}/H)}{3(1+w)} = \psi -\zeta, \eeq using
Eq. (\ref{zeta_l}). Plugging this in Eq. (\ref{SW+ISW}), and using:
\beq \frac{d}{d\eta} = \frac{\partial}{\partial \eta}+
\frac{dr}{d\eta}\hat{\bf n}\cdot\nabla, \eeq to replace partial
derivative with the total (comoving) derivative, we find: \beq
\frac{\delta T(\hat{\bf n})}{T} = -\zeta(r_{LSS}\hat{\bf n}) + 2
\int_0^{r_{LSS}} dr~ \hat{\bf n} \cdot\nabla \psi(r\hat{\bf
  n},\eta_0-r).\label{ISW_grad} \eeq Given that on superhorizon
scales, $\nabla\psi$ is non-vanishing only at the boundaries of the
bubble, we can find a closed form for the Sachs-Wolfe + early ISW
effects: \bea
&&\left[\frac{\delta T(\hat{\bf n})}{T}\right]_{SW+e-ISW} =\nonumber\\ &&\zeta_* \left[-\theta(\eta_{LSS}-\eta_{en})\theta(\eta_{ex}-\eta_{LSS})+ 2g(\eta_{ex})\theta(\eta_{ex}-\eta_{LSS})-2g(\eta_{en})\theta(\eta_{en}-\eta_{LSS})\right],\nonumber\\ \label{eq:SW_eISW}
\eea where $\eta_{en}$ and $\eta_{ex}$ are the conformal times at the
moments that photon enters and exits the bubble respectively (see Figure \ref{fig:LSS}), while
$\theta$ is the Heaviside step function.  Deep into the matter era,
this result reduces to the standard $\zeta_*/5$ for a bubble that
intersects the last scattering surface, and vanishes otherwise.

\begin{figure}
  \centering
  \includegraphics[width=0.75\linewidth]{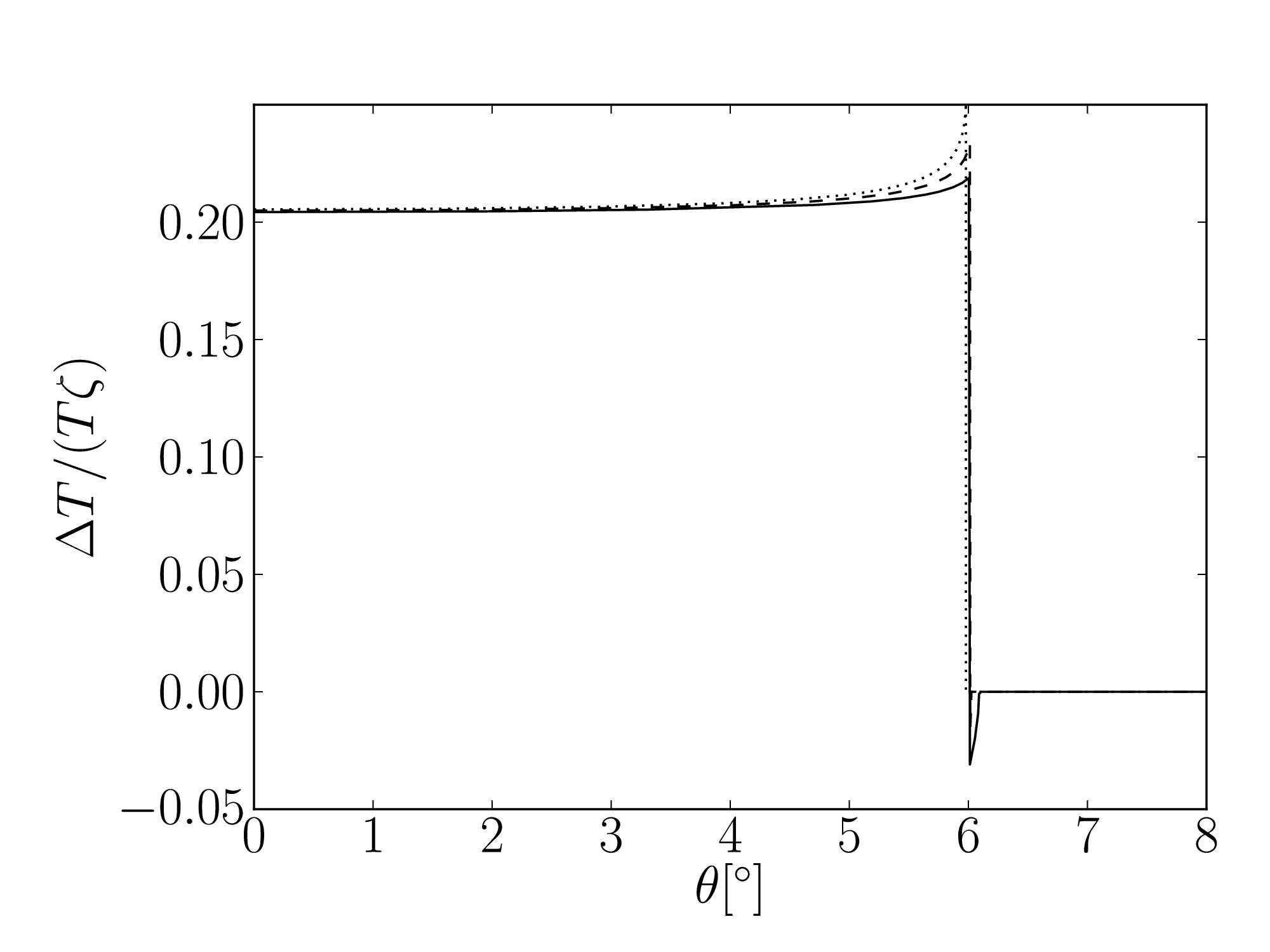}
\caption{This plot shows the profile of a super-horizon bubble of
  $1~ {\rm Gpc}/h$ radius, centered at $z=500,800,1200$ (solid, dashed, dotted)
  lines. At small distances from the center of the bubble, the value of
  $g$ at the photon exit is constant and corresponds to deep matter
  era. At the lines of sight at the edge of the bubble, $g$ is
  increasing. At the very edge of the bubble, photons originate outside
  the bubble for low redshift bubbles and hence the profile reflects the
  difference in the value of $g$ between bubble entry and exit. The edges of this profile will be smoothed on the scale of the sound horizon at last scattering, which is a about a degree.
\label{fig:prof}}

\end{figure}

In Figure \ref{fig:prof}, we plot the profile of such bubble, although note that the edges of this profile will be smoothed on the scale of the sound horizon at last scattering, which is a about a degree. In order to find this profile, we find $\eta_{en}$ and $\eta_{ex}$ for the line of sight at angle $\theta$ from the center of the bubble (Figure \ref{fig:LSS}), use Eq. (\ref{eq:g_analytic}) to find $g(\eta_{ex})$ and $g(\eta_{en})$, and plug into Eq. (\ref{eq:SW_eISW}) to find $\delta T/T$.

\subsection{Subhorizon bubbles: Late-time ISW and Rees-Sciama effects}

\begin{figure}
  \centering
  \includegraphics[width=0.45\linewidth]{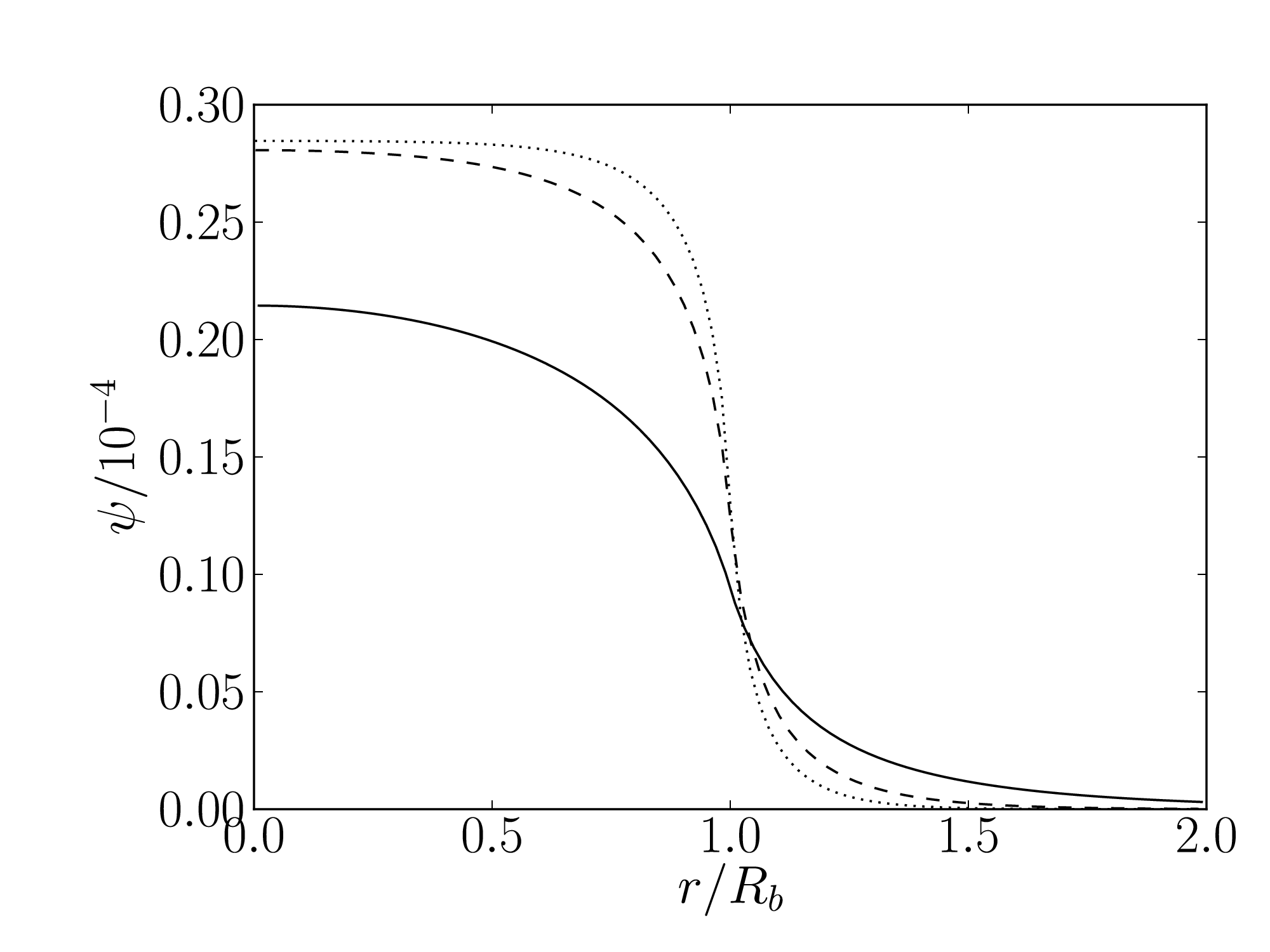}
  \includegraphics[width=0.45\linewidth]{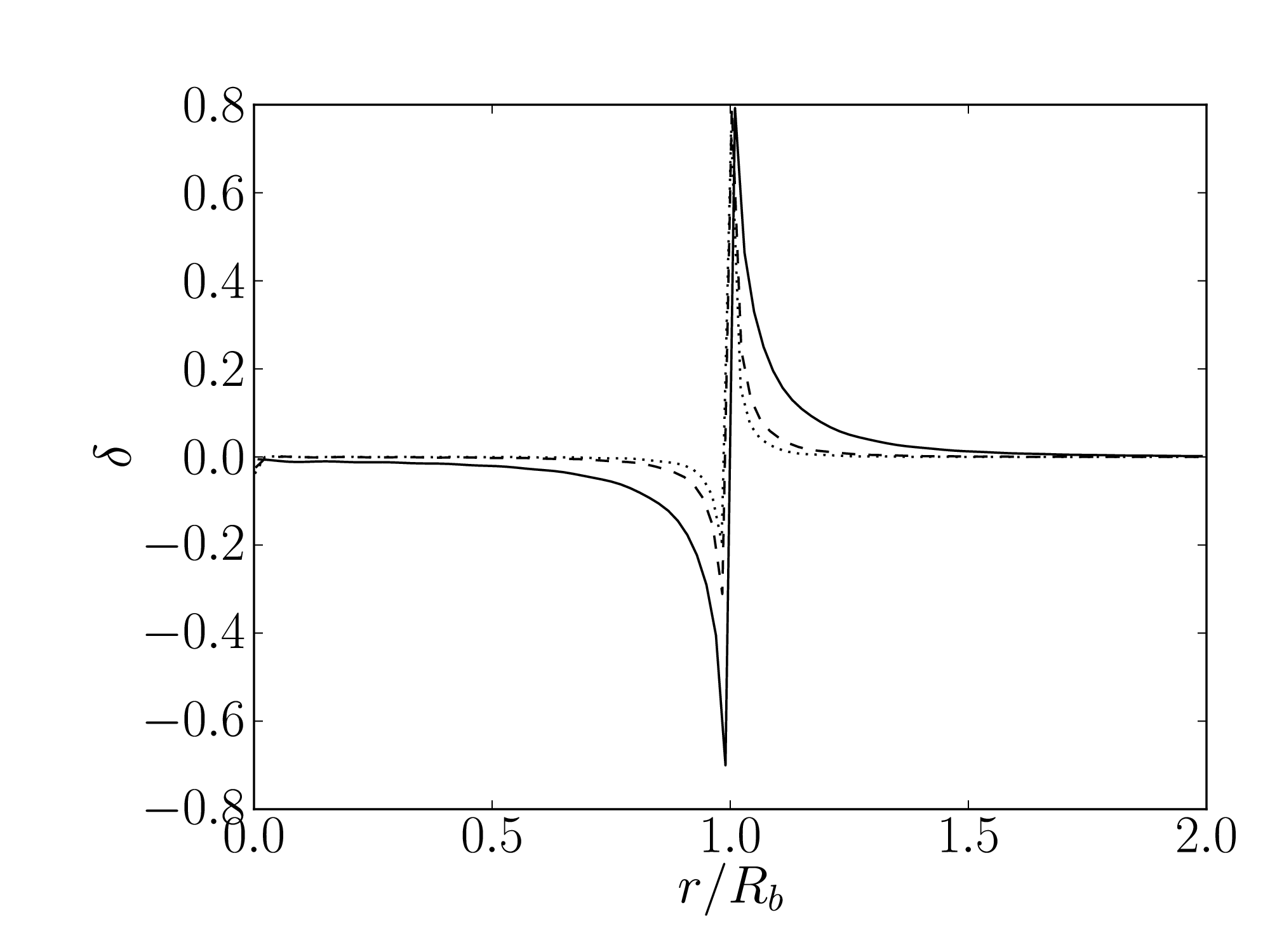}
  \caption{This figure shows the bubble profile in potential (left
    panel) and overdensity (right panel) for $\zeta_*=5\times
    10^{-5}$ for buble size of 100 Mpc/$h$ (solid), 300 Mpc/$h$
    (dashed) and 500 Mpc/$h$ (dotted) at redshift $z=1$. Note that
    bubbles considered in this paper have the same mean density inside
    the bubble as the mean cosmic density.}
  \label{fig:psidelta}
\end{figure}

For bubbles that had sub-horizon size during the radiation dominated
era, the boundaries will not be very sharp as the primordial top-hat
profile should be convolved with the transfer function
$T(k)$. Assuming linear evolution, and using Eq. (\ref{psi_k}), this
implies: \bea
\psi({\bf r},\eta) = g(t) \zeta_* S(|{\bf r}-{\bf r}_c|),\\
S(|{\bf x}|) = \frac{4\pi R^3_{\rm bubble}}{3}\int \frac{d^3{\bf
    k}}{(2\pi)^3} \exp(i{\bf k}\cdot {\bf x}) T(|{\bf k}|) W(|{\bf
  k}|R_{\rm bubble}), \eea where \beq W(x) = 3x^{-3}(\sin x - x \cos x),
\eeq is the top-hat filter in the Fourier space. We illustrate this in the Figure \ref{fig:psidelta}.

To find the linear late-time ISW effect \cite{Sachs:1967er}, it is
sufficient to assume the bubble profile does not change in a light
crossing time. Within this approximation, the impact on the CMB just
depends on the projection of the bubble profile $S(r)$: \bea
\left[\frac{\delta T(b)}{T}\right]_{lISW} = 2 \int d\eta
\frac{\partial \psi}{\partial \eta} \simeq 2a \dot{g}\zeta_*
\int^{\infty}_{-\infty} dy ~S\left(\sqrt{b^2+y^2}\right) \nonumber\\=
\frac{8\pi a\dot{g} \zeta_* R^3_{\rm bubble}}{3}\int \frac{d^2{\bf
    k}}{(2\pi)^2} \exp(i{\bf k}\cdot {\bf b}) T(|{\bf k}|) W(|{\bf
  k}|R_{\rm bubble}), \eea where $b$ is the distance by which the line
of sight misses the center of the bubble.

Since the late-time (linear)
ISW effect depends on $\dot{g}$, it only becomes significant when dark
energy becomes important. Therefore, since dark energy is sub-dominant
at high redshifts, an over/underdensity in the matter era can only
contribute to ISW effect at the non-linear level. This is known as the
Rees-Sciama effect \cite{1968Natur.217..511R}.

In order to find the Rees-Sciama effect, we can use the $G_{ij}$
Einstein equations, deep in the matter era, which take the form: \beq
\left[\psi''+3aH\psi'+\frac{1}{2}\nabla^2(\phi-\psi)\right]\delta_{ij}-\frac{1}{2}(\phi-\psi)_{,ij}
\simeq -4\pi G T^i_j = 4\pi G \rho_m a^2 u^i u^j,\label{G_ij} \eeq
where $' \equiv \partial_{\eta}$, while $u^i$'s are the components of the peculiar velocity, and can be
fixed using the matter-era $G_{0i}$ constraint: \beq -H \psi_{,i} =
4\pi G\rho_m a u^i. \label{G_0i} \eeq For spherically symmetric scalar
fields $X$, \beq X_{,ij}=\frac{\partial}{\partial
  x^j}\left(\frac{\partial X}{\partial r} \frac{\partial r}{\partial
    x^i}\right) = \frac{x^i x^j}{r} \frac{\partial}{\partial
  r}\left[r^{-1}\frac{\partial X}{\partial r}\right]\eeq and hence
Eq. (\ref{G_0i}) can be used to write the off-diagonal part of
(\ref{G_ij}) as: \beq r\frac{\partial}{\partial
  r}\left[r^{-1}\frac{\partial (\phi-\psi)}{\partial r}\right] = -
\frac{4}{3}(\nabla\psi)^2,\label{G_ij_off} \eeq while its trace takes
the form \beq 3(\psi''+3aH\psi')+r^{-2}\frac{\partial}{\partial
  r}\left[r^{2}\frac{\partial (\phi-\psi)}{\partial r}\right] =
\frac{2}{3}(\nabla\psi)^2. \label{G_ii} \eeq Equation (\ref{G_ij_off})
can be integrated using 1st order solution for $\psi$ and inserted
into (\ref{G_ii}), yielding an equation for time evolution of $\psi'$
at second order. For fixed $g(t)=3/5$ in the matter era, this can be
solved to give \beq \psi'^{(2)}({\bf r},\eta) = \frac{18\zeta^2_*}{175
  a H}U(|{\bf r}-{\bf r_c}|), \eeq where \beq U(r) \equiv \frac{2}{3}
\left[dS(r) \over dr\right]^2 -\frac{4}{3} \int_r^{\infty}
\frac{dr'}{r'} \left[dS(r') \over dr'\right]^2, \eeq and we have
ignored the transient homogeneous solution to
Eq. (\ref{G_ii}). Similar to ISW effect, the Rees-Sciama effect is the
line of sight integral of $\psi'$: \beq \left[\frac{\delta
    T(b)}{T}\right]_{RS} = 2 \int d\eta ~\psi'^{(2)} =
\frac{36\zeta^2_*}{175 a H} \int^{\infty}_{-\infty} dy
~U\left(\sqrt{b^2+y^2}\right), \eeq where we have again assumed that
the bubble light crossing time is much shorter than the Hubble time.
As expected, the total effect is dominated by the ISW contribution at
low redshifts and by the Rees-Sciama at higher redshifts.

\subsection{Subhorizon bubbles: Ostriker-Vishniac effect}

Another secondary anisotropy that could be produced by large bubbles
is the Ostriker-Vishniac (or the kinetic Sunyaev-Zel'dovich effect)
\cite{1986ApJ...306L..51O}. This effect is caused by the Doppler shift
of the CMB photons, scattered by free electrons in the universe. It is
given by: \beq \left[\frac{\delta T(\hat{\bf n})}{T}\right]_{OV} =
-\int_0^{r_{LSS}} dr \frac{d\tau}{dr} \left[1+\delta_e(r\hat{\bf
    n},\eta_0-r) \right] \left[\hat{\bf n}\cdot {\bf u}(r\hat{\bf
    n},\eta_0-r)\right],\label{OV} \eeq where $\delta_e$ is the
overdensity of free electrons, and $d\tau/dr$ is the cosmic mean
differential optical depth for Thomson scattering: \beq d\tau =
\bar{n}_e a \sigma_T dr.  \eeq Here, $\bar{n}_e$ is the mean physical
number density of free electrons in the universe, while $\sigma_T$ is
Thomson cross-section.

On the scales that have not gone through shell-crossing, and assuming
a uniform ionization ratio, the electron and total matter
overdensities are the same: $\delta_e \simeq \delta_m$. Therefore,
$G_{0i}$ Einstein equation: \beq \frac{(ag)'}{g} \psi_{,i} = -4\pi G
a^3\rho_m (1+\delta_m) u^i, \eeq can be used to fix the kernel in
Eq. (\ref{OV}). After integration by part, this yields: \bea
\left[\frac{\delta T(\hat{\bf n})}{T}\right]_{OV} = \frac{f_b(1+X)\sigma_T}{8\pi G m_p} \int_0^{r_{LSS}} dr  g^{-1} \psi(r \hat{\bf n},\eta) \frac{\partial}{\partial \eta}\left[\frac{x_e}{a^2} \frac{\partial (ag)}{\partial \eta}\right]\nonumber\\
\simeq\frac{f_b(1+X)\zeta_*\sigma_T}{8\pi G m_p
  }\left\{\frac{\partial}{\partial \eta}\left[\frac{x_e}{a^2}
    \frac{\partial (ag)}{\partial \eta}\right] \right\}
\int^\infty_{-\infty} dy~ S\left(\sqrt{b^2+y^2}\right) \nonumber \\
= \frac{3 f_b(1+X)\zeta_*\sigma_T H a^2}{16\pi G m_p}\left\{\frac{\partial}{\partial
    a}\left[x_eH(1-g)(1+w)\right]\right\} \int^\infty_{-\infty} dy~
S\left(\sqrt{b^2+y^2}\right), \label{eq:ov}\eea where $m_p$ is the proton mass, $X
\simeq 0.75$ is the Hydrogen mass fraction in the intragalactic
medium, and $x_e$ is the ionized fraction (assuming both hydrogen and
helium are ionized equally) and $f_b=\Omega_b/\Omega_m$ is the baryon
fraction. $H$ and $w$ are Hubble constant and total equation of state (total pressure divided by total density), respectively. 

\begin{figure}
  \centering
  \includegraphics[width=0.75\linewidth]{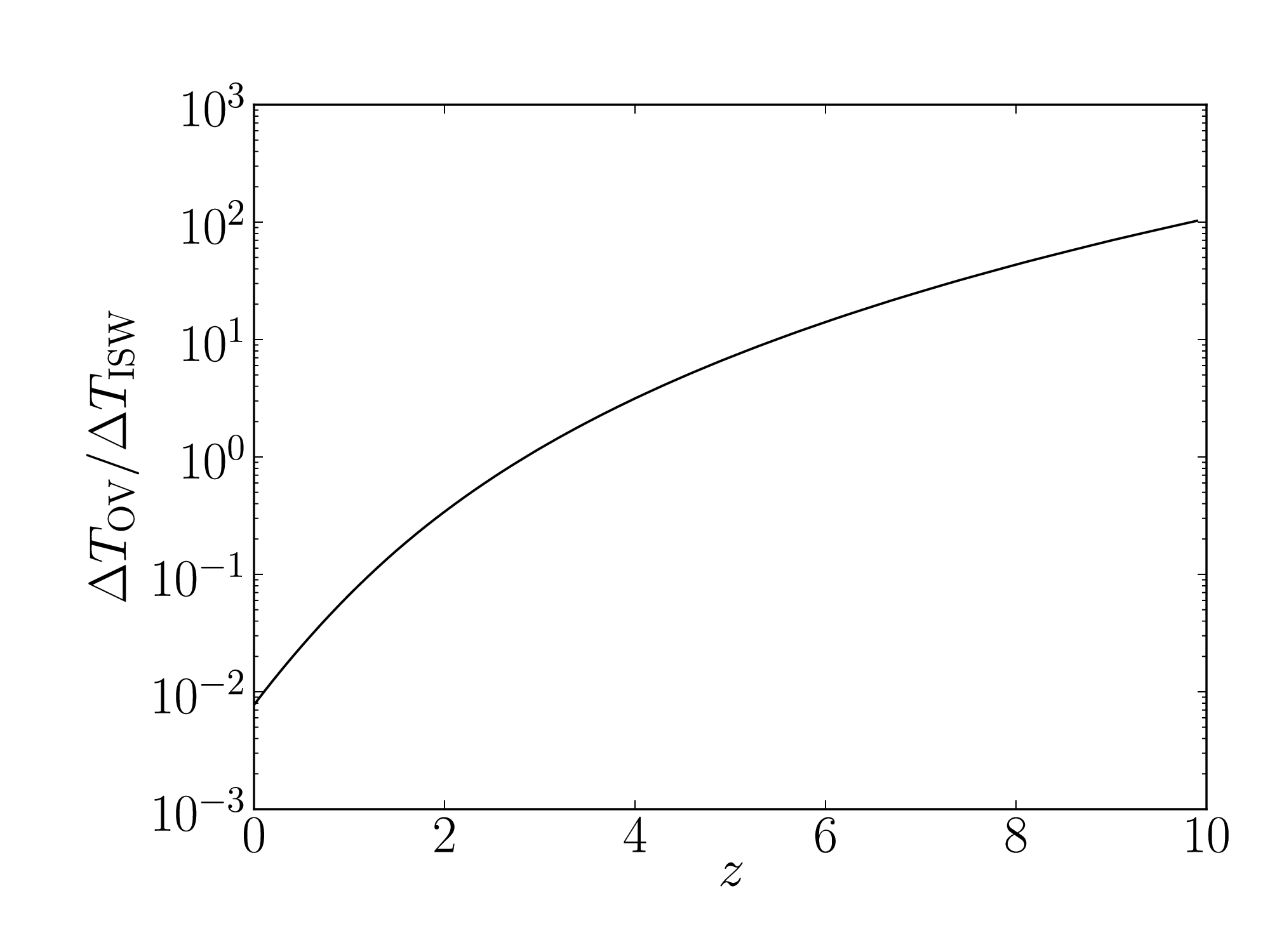}
  \caption{This figure shows the relative amplitudes of
    Ostriker-Vishiniac and ISW effects as a function of redshift. We
    assume $X=0.75$ for hydrogen abundance, and fully ionized
    medium. We ignore the fact that Helium was most likely neutral at
    redshift bigger than $\sim3$.}
  \label{fig:gofa}
\end{figure}

Since both ISW and OV effects scale as $\zeta_* \int S$, it is easiest to
asses the relative strength of the two effect by plotting the ratio of
relevant quantities. We do this in the Figure \ref{fig:gofa}. We note
that at redshifts of $\sim 3$, the two effects become comparable and that for
larger redshifts, the OV begins to dominate.

\begin{figure}
  \centering
  \begin{tabular}{lll}
  \includegraphics[width=0.3\linewidth]{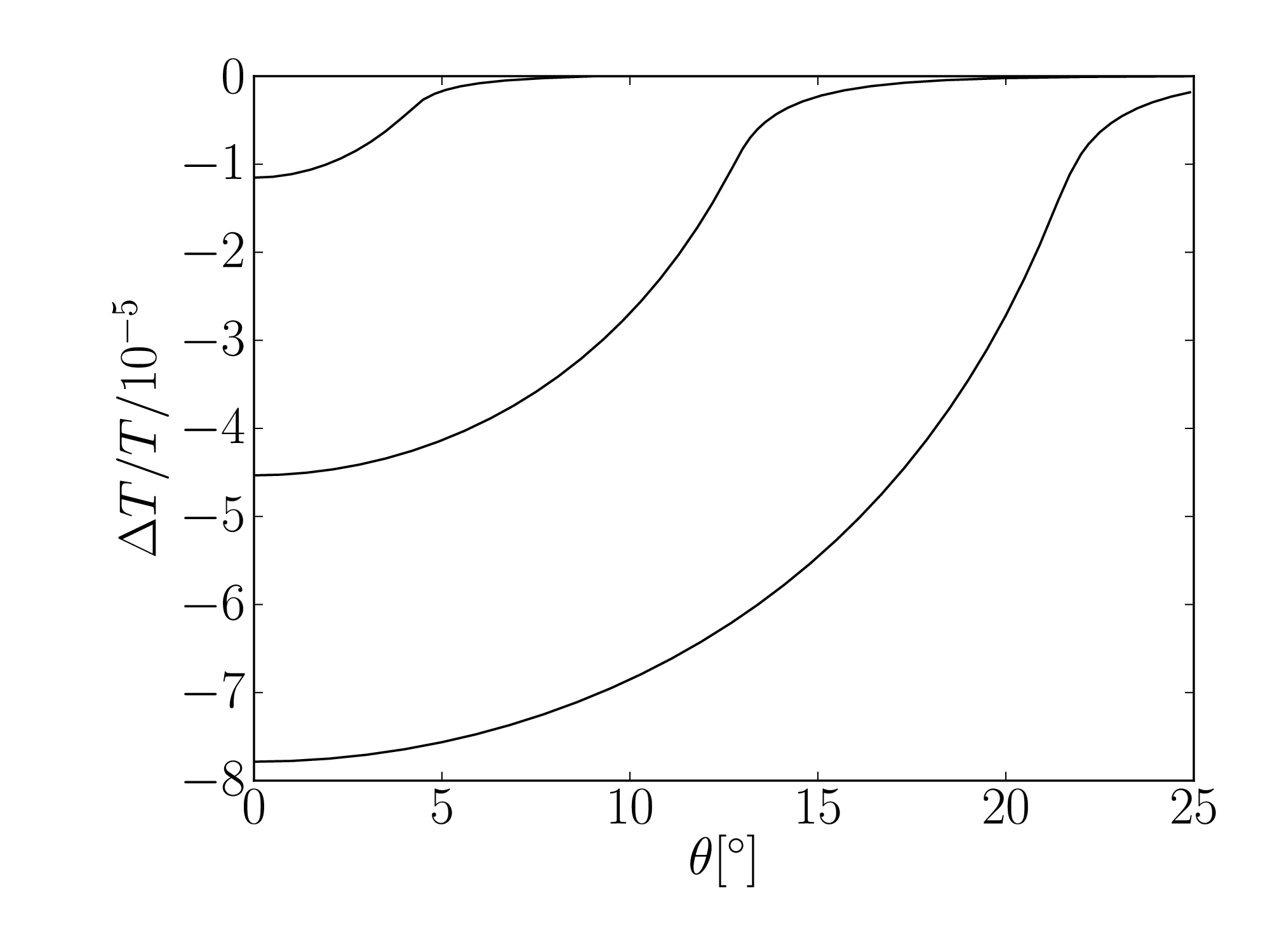} &
  \includegraphics[width=0.3\linewidth]{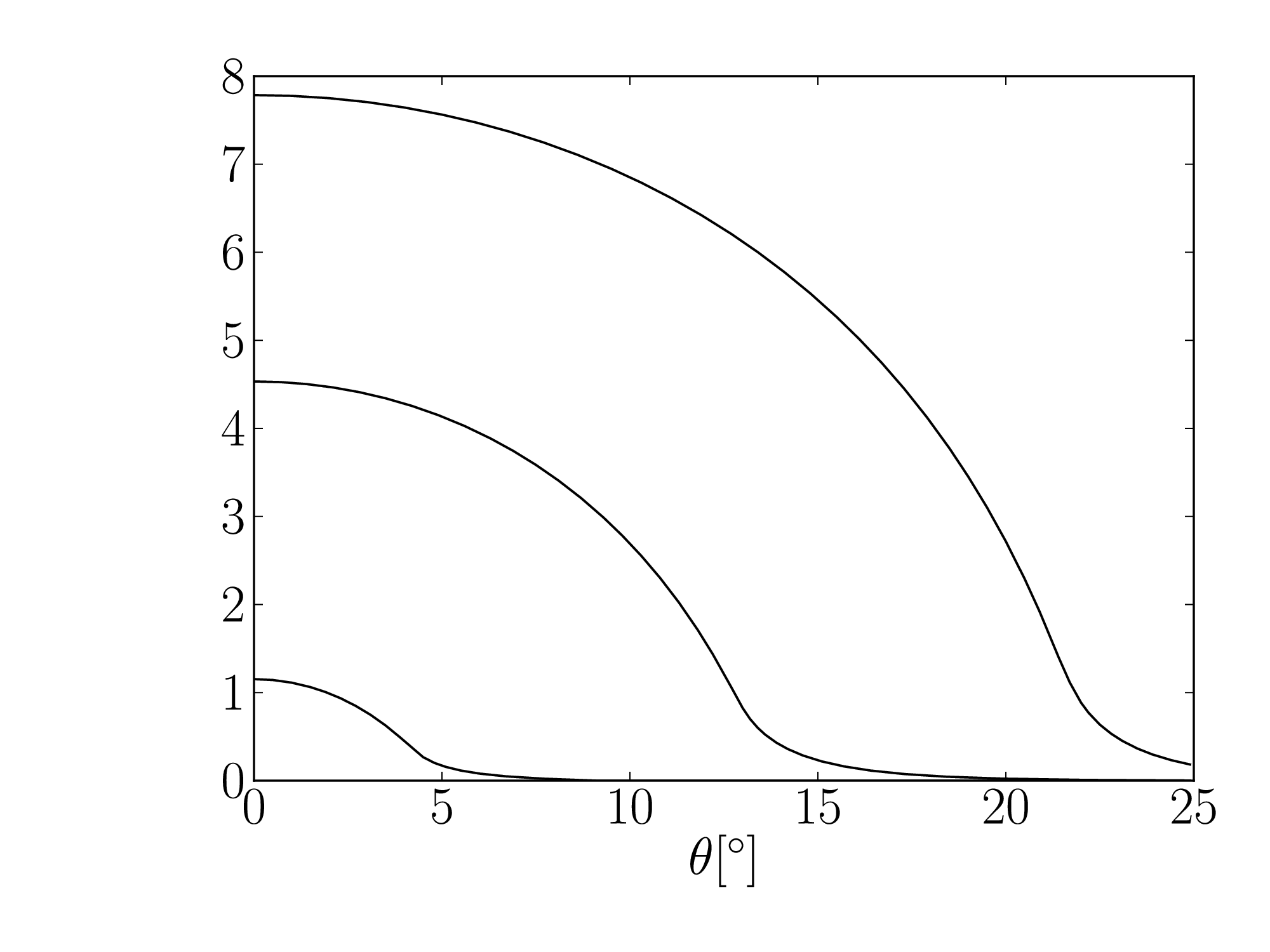} &
  \includegraphics[width=0.3\linewidth]{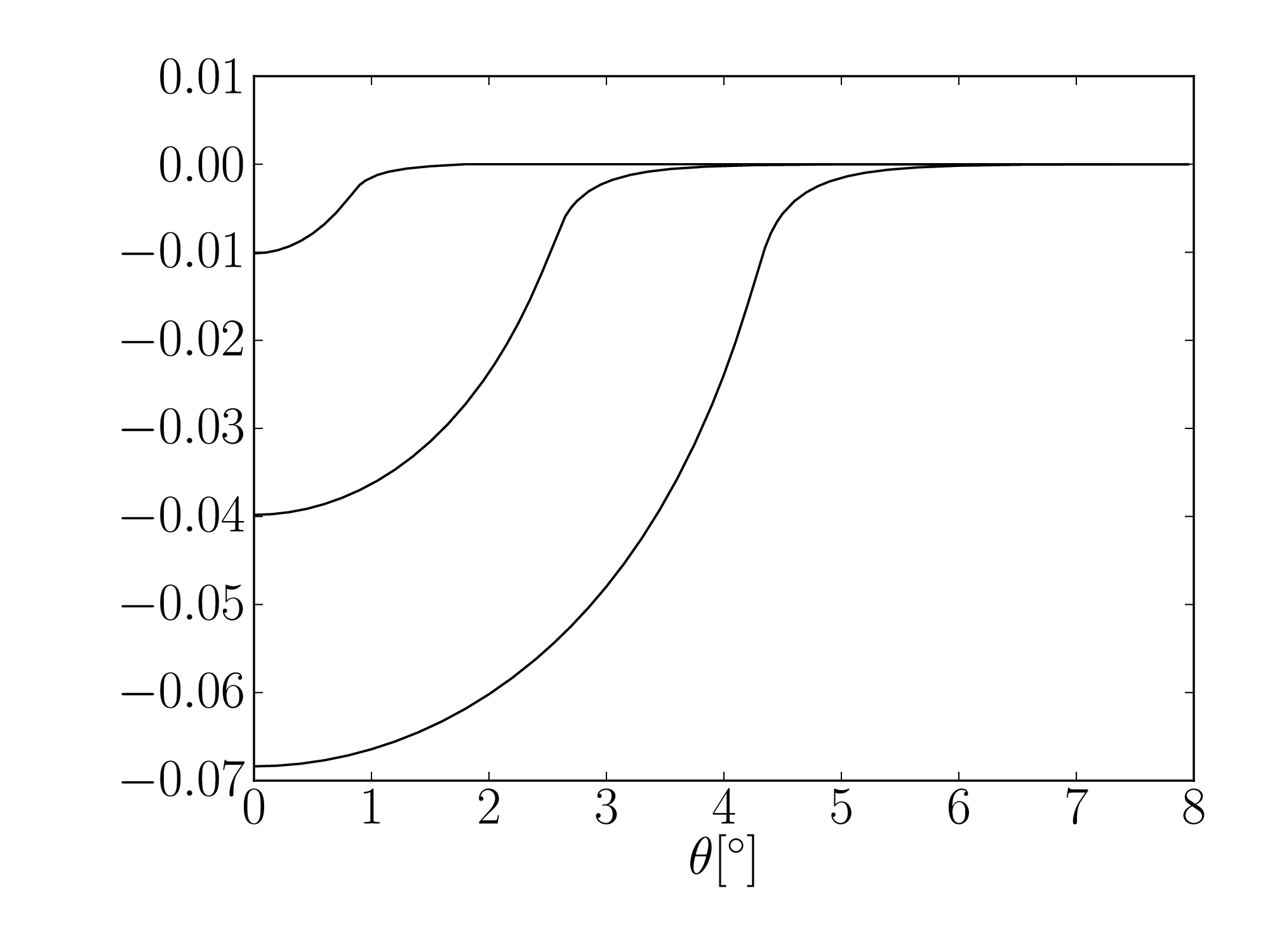} \\
  \includegraphics[width=0.3\linewidth]{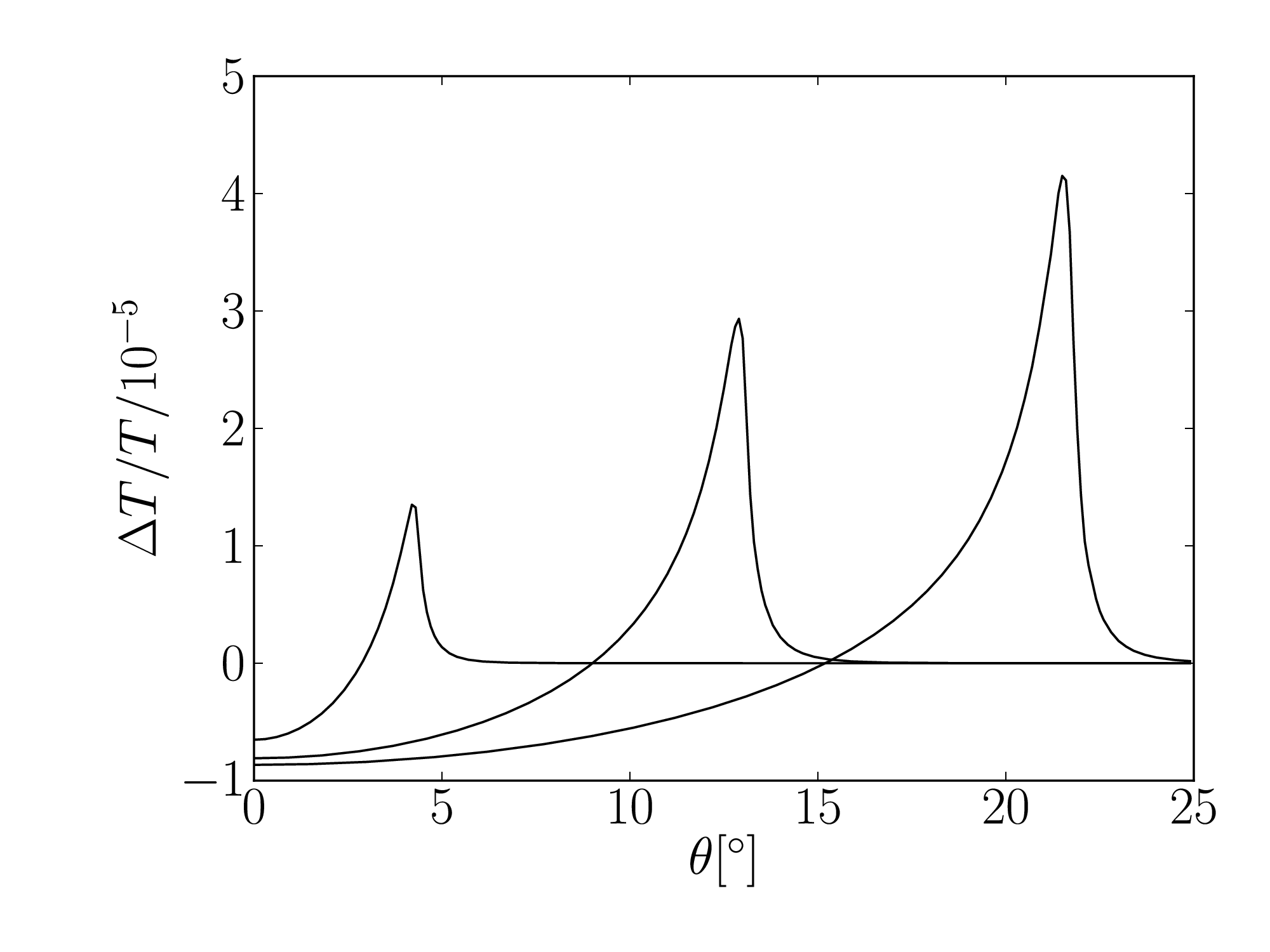} &
  \includegraphics[width=0.3\linewidth]{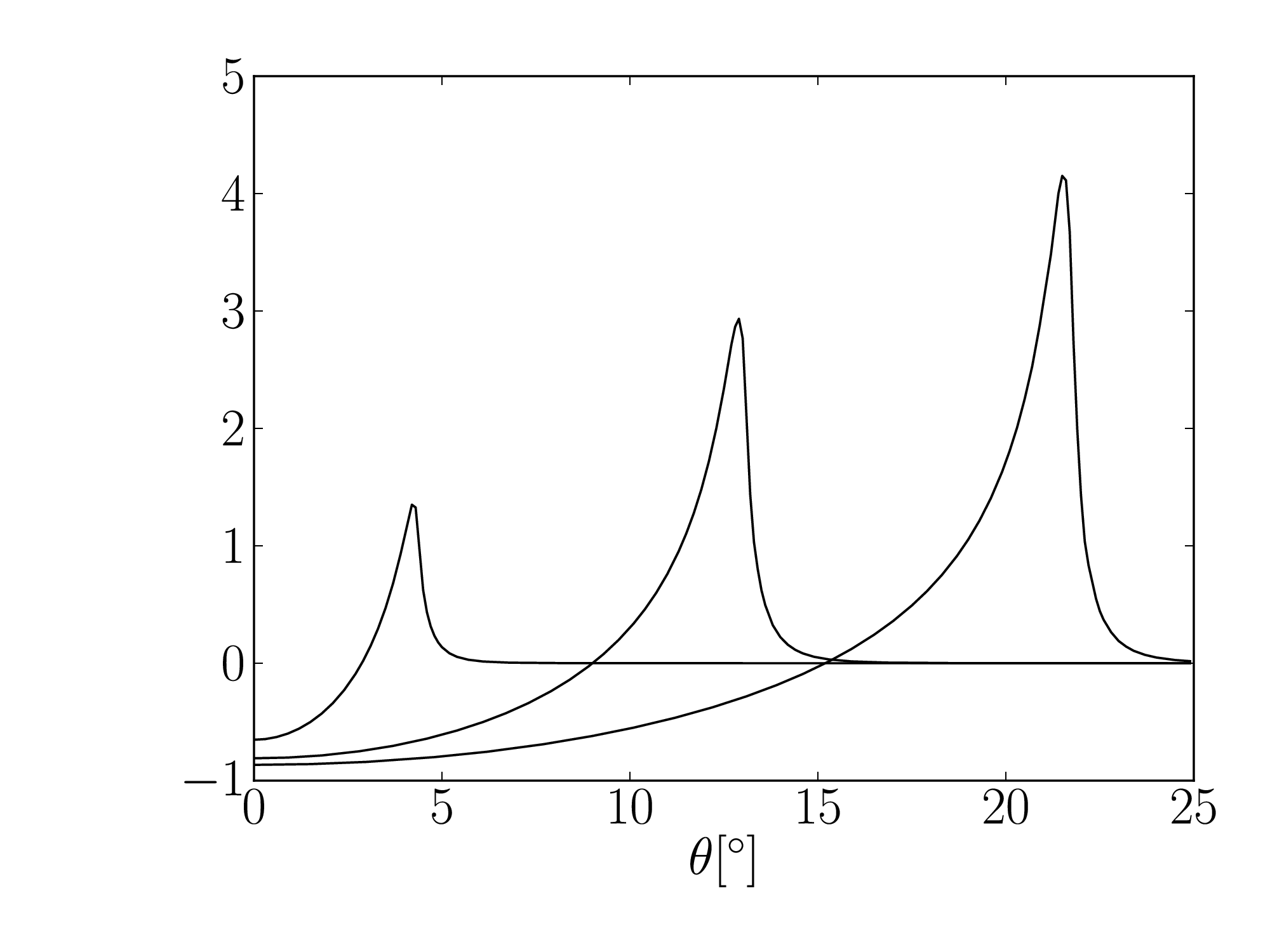} &
  \includegraphics[width=0.3\linewidth]{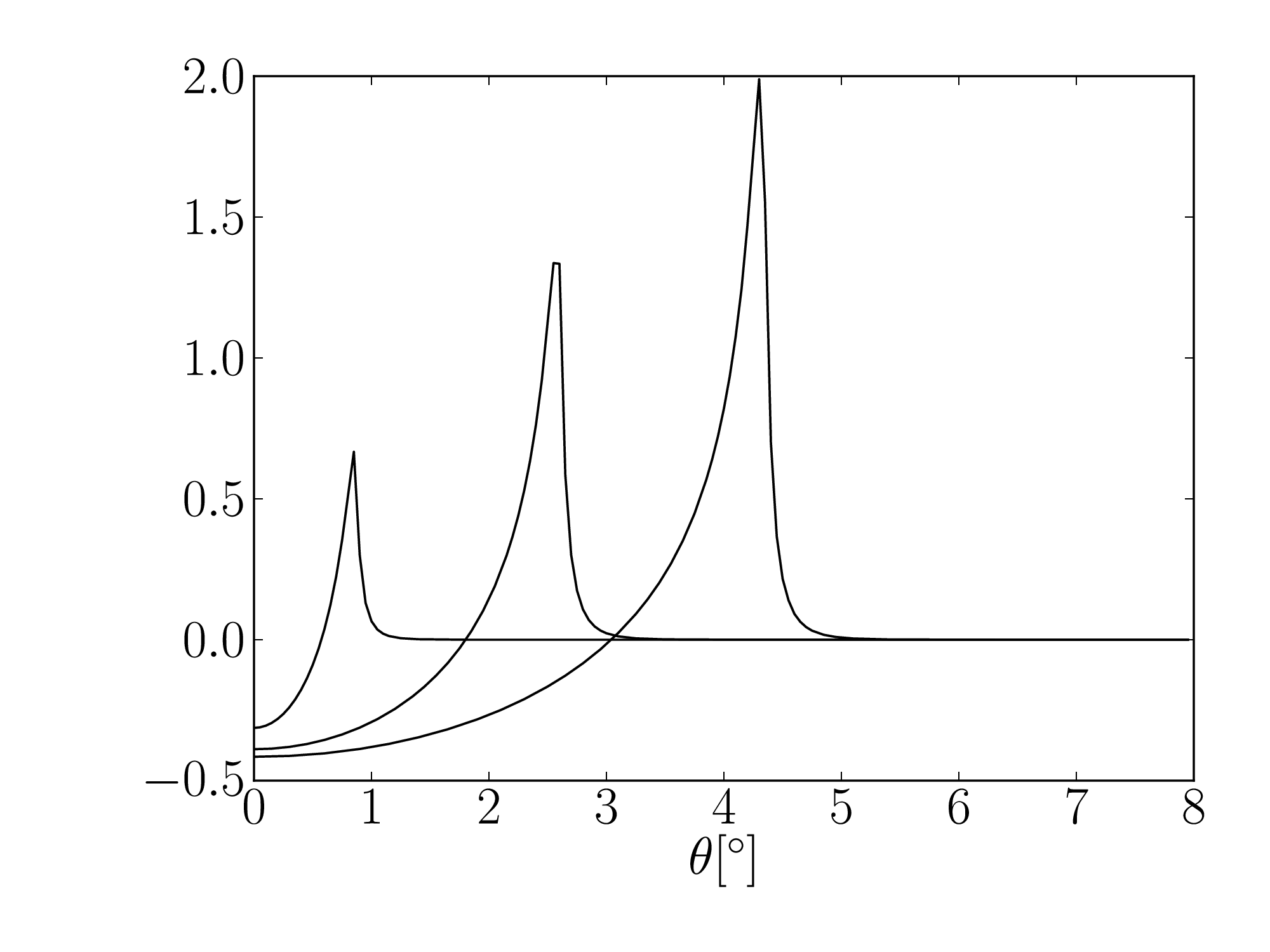} \\
  \includegraphics[width=0.3\linewidth]{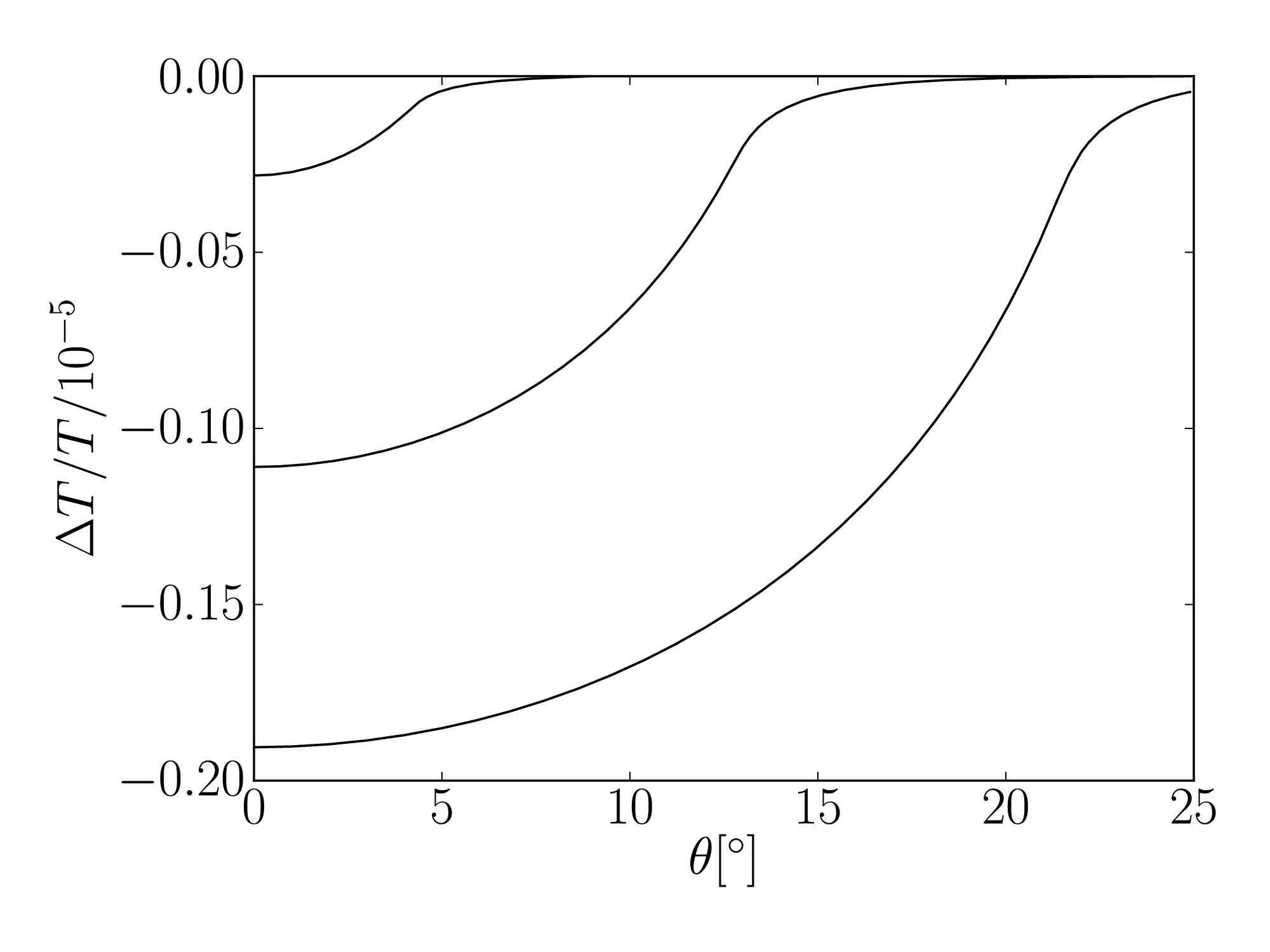} &
  \includegraphics[width=0.3\linewidth]{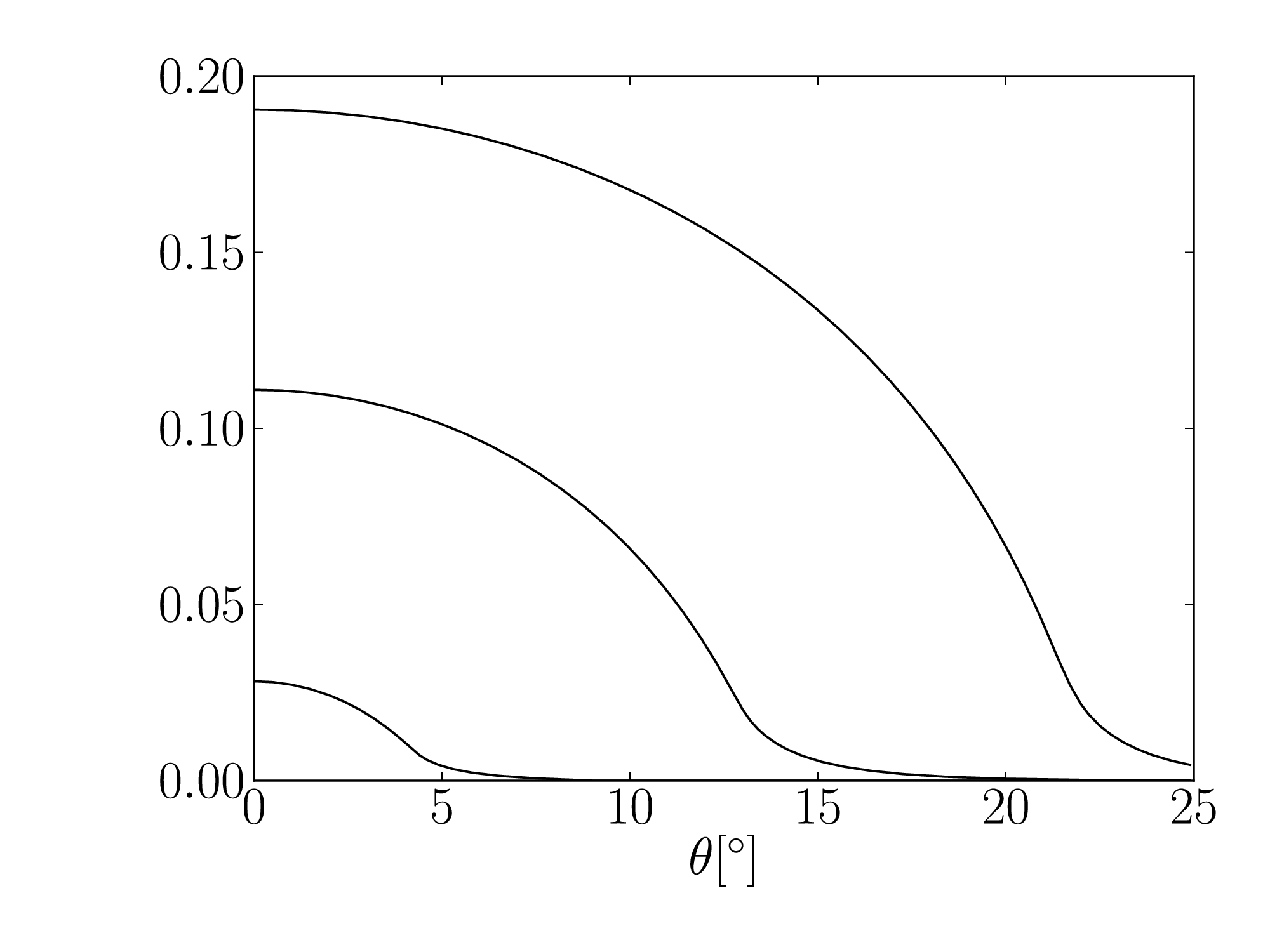} &
  \includegraphics[width=0.3\linewidth]{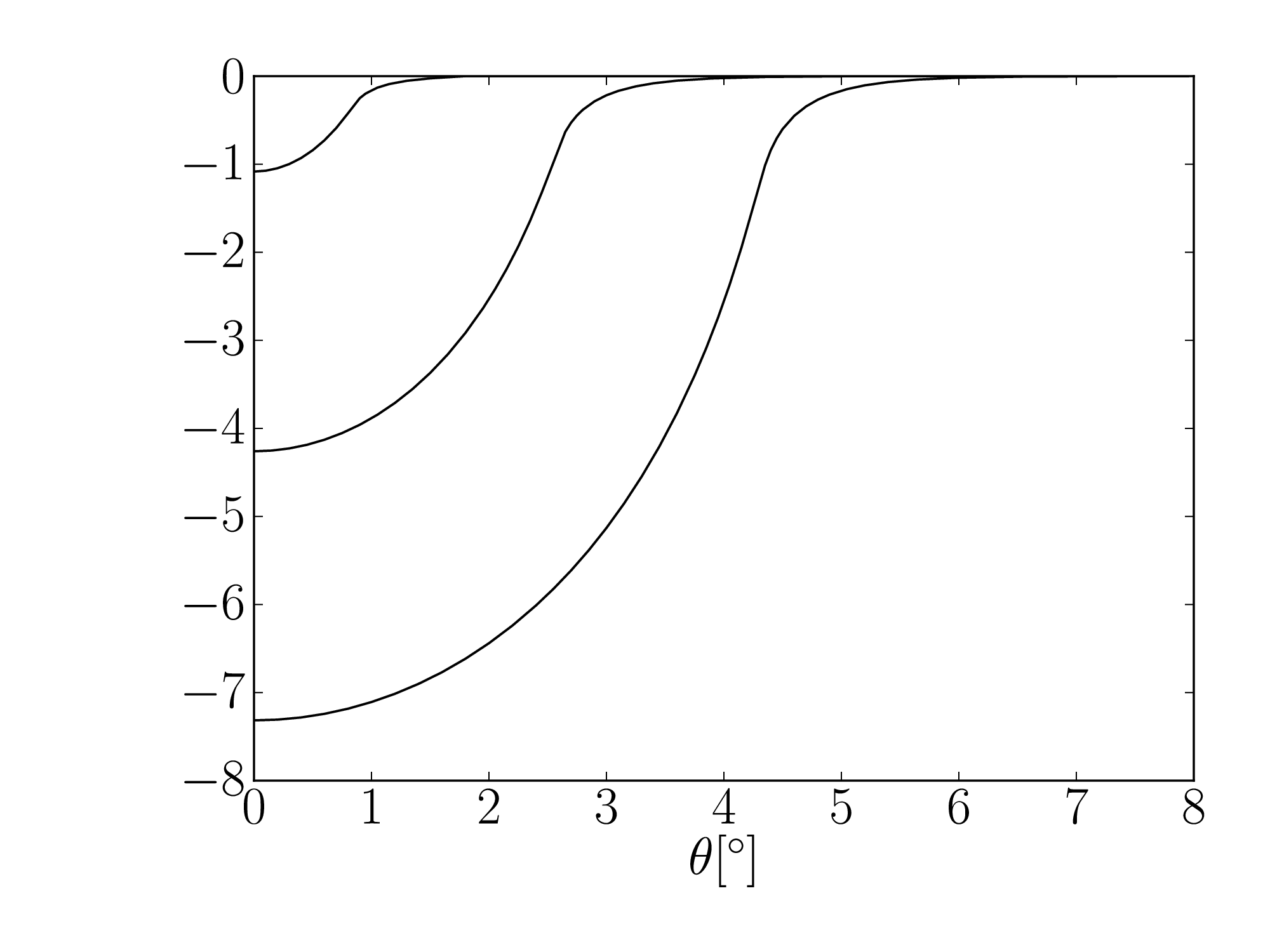} \\
  \includegraphics[width=0.3\linewidth]{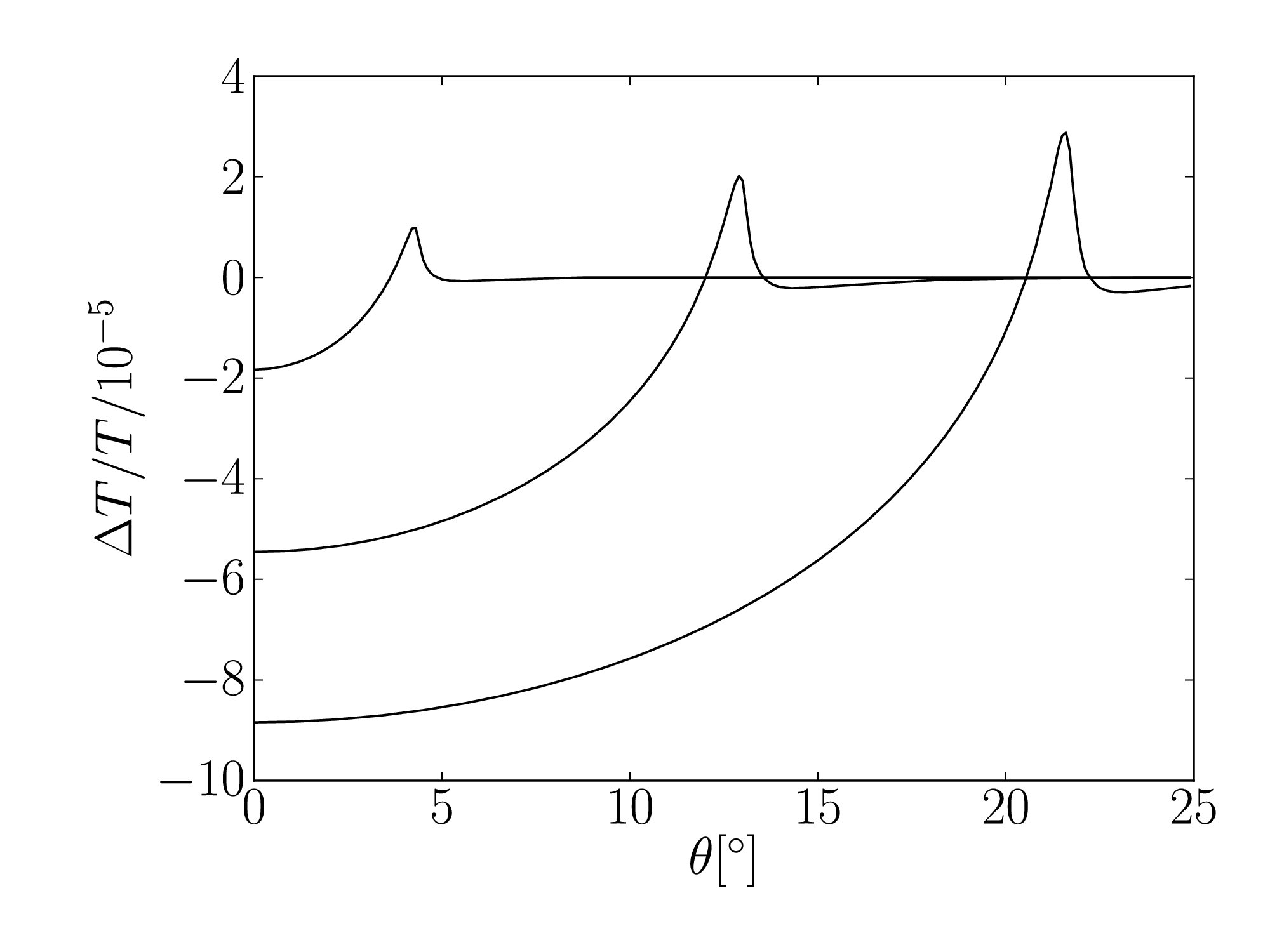} &
  \includegraphics[width=0.3\linewidth]{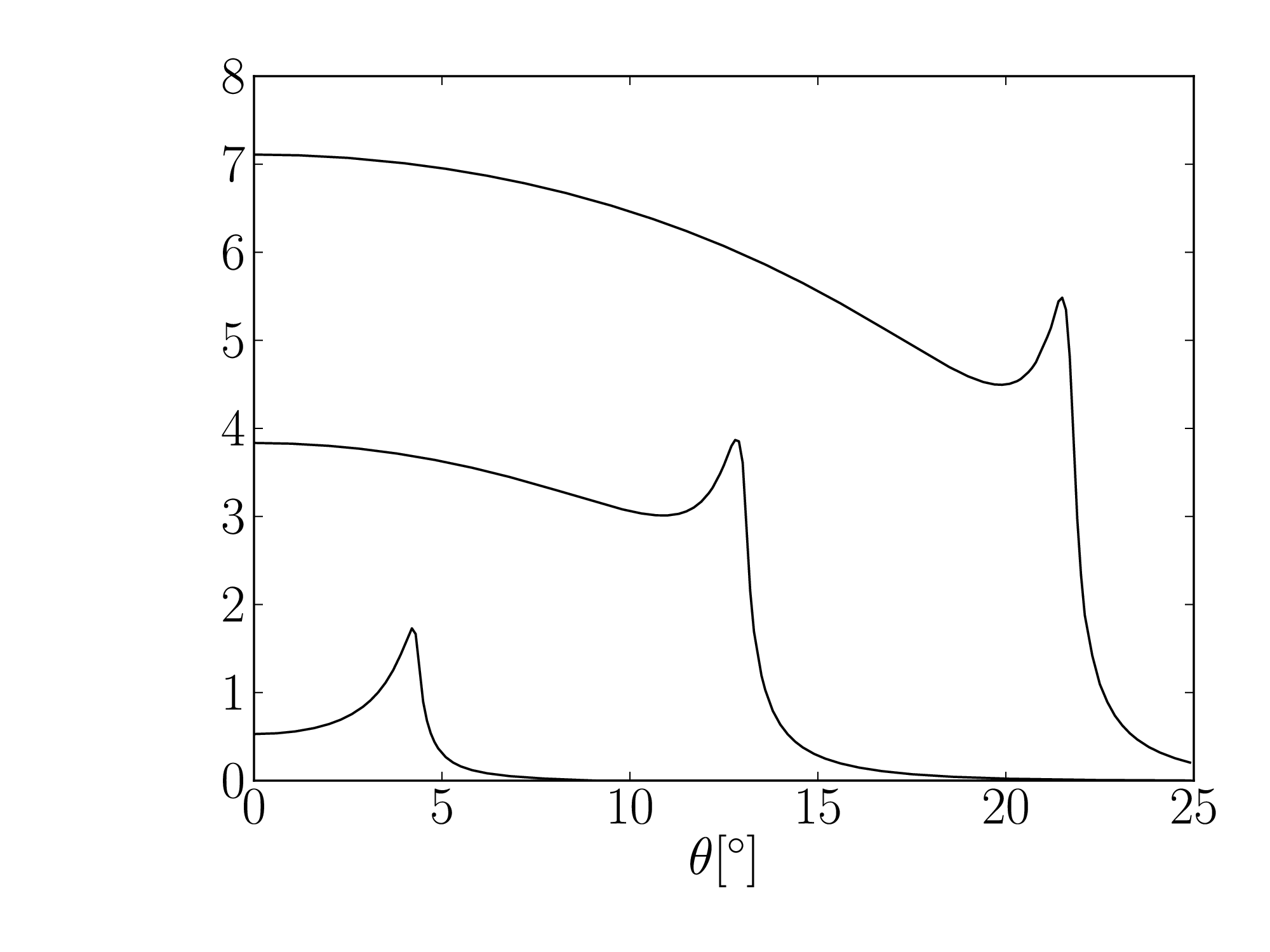} &
  \includegraphics[width=0.3\linewidth]{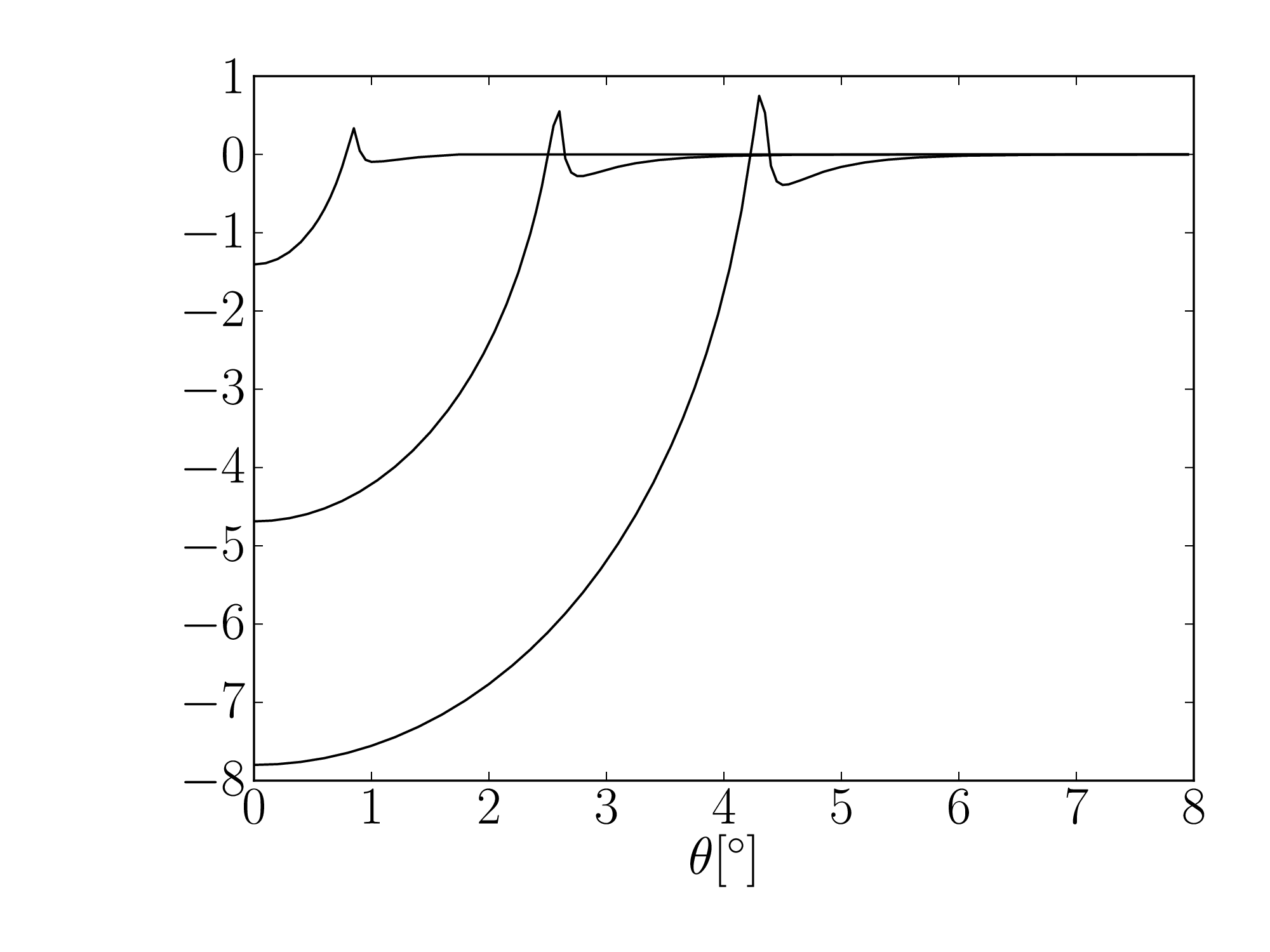} \\
  \end{tabular}
  \caption{This figure shows the bubble profile associated with ISW
    (topmost row), Rees-Sciama (second row),
    Ostriker-Vishiniac (third row) and total contribution
    (bottom row). We show the profile for bubbles size 100, 300 and 500 Mpc/$h$
    at redshift $z=0.5$ with $\zeta_*=10^{-3}$ (left panels) and
    $\zeta_*=-10^{-3}$ (middle panels) and at redshift
    $z=10$ with $\zeta_*=10^{-3}$ (right panels).     
    \label{fig:prof22}}
\end{figure}

We plot the individual effects discussed here for three representative
cases in the Figure \ref{fig:prof22}. The $\zeta$ perturbations were
chosen to give a perturbation of $\delta T/T\sim 10^{-4}$ which is the
same order of magnitude as the cold-spot perturbation. There are
several interesting aspects to note regarding these plots. First, the
ISW effect dominates at small distances, but never completely
overtakes the RS effect for sensible values of $\zeta$ that can give
the right magnitude. At the same time, the sign of $\zeta$ determines
whether the two effects are going to interfere constructively or
destructively. Second, the ISW and OV effects dominate at small and
large redshift respectively and conspire to give similar sized
effects. The result is that the total effect has a non-trivial angular
structure and redshift dependence and this work provides clear
templates that can be tested against real CMB data.

It is interesting to note that a sharp change in the ionization fraction $x_e$, e.g. at reionization, could also lead to a significant OV effect in Eq. (\ref{eq:ov}), as was first pointed out in \cite{Alvarez:2005}.

Note that, for the purpose of calculation of RS effect in Figure \ref{fig:prof22}, we assumed the Universe is still in the matter domination at $z=0.5$. Given that matter domination ends around $z\sim 0.5$, we expect this assumption to introduce $O(50\%)$ error in RS, and $O(10-20\%)$ error in total temperature profile.

\section{Comparison with Cold Spot}\label{sec:comp-cold-spot}
The mechanism proposed in this section can, in principle, produce bubble
regions that can be larger or smaller than the horizon. Therefore, we
can explain the bubble using two different scenarios: bubble between us
and the surface of the last scattering and a super-horizon
overdensity at the actual surface of the last scattering.

If a bubble close to or intersecting the surface of last scattering is responsible for the cold spot, the 4 degree angular radius of the cold spot requires a comoving radius of $R_{\rm bubble} \sim 1~ {\rm Gpc}$. Figure (\ref{fig:prof}) shows that a temperature decrement of $70~ \mu K = 3 \times 10^{-5} T_{\rm CMB}$ requires $\zeta_* = -1 \times 10^{-4}$. Furthermore, given that there must be one bubble within a Gpc of the last scattering surface (at radius of $\sim 10$ Gpc), one expect $N_{\rm observable} \sim 4$ bubbles within our observable horizon. Plugging these into inflationary scenario of Sec. \ref{sec:bubble-inflation} gives:
\beq
p \sim 0.15, ~\Delta N_1^* \sim 1, ~\Delta N_2 \simeq -1 \times 10^{-4},~~{\rm for~superhorizon~bubbles}.
\eeq

For bubbles at progressively lower redshifts, the radius of bubble should be smaller to match the observed extent of the cold spot. Given the scarcity of anomalies such as cold spot, we do not expect the bubble to be closer to us than $z\sim 1$, assuming that we sit at a random point in the universe. This gives a minimum radius of $\sim 200$ Mpc for the bubble, which implies:
\beq
1  \lesssim \Delta N_1^* \lesssim 4,
\eeq
from Eq. (\ref{eq:r_bubble}), given that $\Delta N_2 = \zeta_* \ll 1$.

Interestingly, we can also estimate $p$, simply based on the angular size of the bubble. To do this, we first note that the angular size of the cold spot implies a comoving radius of:
\beq
R_{\rm bubble} \sim (1 ~{\rm Gpc}) (r/r_{LSS}),
\eeq
if the bubble sits at the comoving distance $r$ from us. However, if the closest (and thus easiest to see) bubble is at distance $r$, there should be
\beq
N_{\rm observable} \sim (r_{LSS}/r)^3,
\eeq
bubbles within our observable horizon. Now, combining Eqs. (\ref{eq:n_obs}) and (\ref{eq:r_bubble}) yields:
\beq
p \sim N_{\rm observable} \left(H_0 R_{\rm bubble}\right)^3 \sim 0.04,
\eeq
given that $H^{-1}_0 = 3.0~{\rm Gpc}/h$. In other words, bifurcation into path $B$ is roughly a $2\sigma$ event in the inflationary history.

Figures (\ref{fig:prof22}) and (\ref{fig:gofa}) show that in order to match a temperature decrement of $\sim 10^{-5}$ for bubbles at $0.5< z <10$:
\beq
\zeta_* = \Delta N_2 \sim 10^{-3},
\eeq
i.e. the bifurcated inflationary path (path B) expanded $0.1$\% more that the regular inflationary trajectory (path A).

%
%

Finally, we should note that in what came above we have ignored the impact of regular inflationary gaussian fluctuations on our predictions. The fact is that both Rees-Sciama and Ostriker-Vishniac effects include 2nd order cross-terms with product bubble profile and standard gaussian fluctuations. While these terms have zero mean, they lead to a random component in the bubble profile, which can be potentially seen in higher order statistics of the CMB sky (e.g., see \cite{Masina:2008zv,Masina:2009wt}). Such treatment is clearly beyond the scope of the current paper, but will be possible given the formulae provided in Sec. (\ref{sec:comp-with-observ}).

\section{Conclusions and Future Prospects}
\label{sec:conclusions}

In this paper we have introduced a novel mechanism that could lead to
over/underdense spherical regions/bubbles in the universe, through a
probabilistic bifurcation in the inflationary trajectory. We have then
worked out predictions for the imprint of such bubbles on the CMB, and
estimated approximate values of the parameters of the inflationary
scenario that could lead to the observed WMAP cold spot.

If our mechanism is correct and the cold-spot observed in the WMAP
data is indeed a massive high redshift bubble, there are several
observational probes that can test this assumption:

Of course, the most immediate test will be to investigate how well the
CMB profiles of Sec. (\ref{sec:comp-with-observ}) can fit the WMAP
cold spot.  Another promising direction is to reconstruct the bubble
profile by detecting the lensing of the CMB by the large bubble
\cite{Das:2008es}, and compare it to our prediction for density
profile (Fig. \ref{fig:psidelta}). Such tests are well within realms
of experiments such as Atacama Cosmology and South Pole Telescopes
\cite{ACT:2010cy,Lueker:2009rx}. Other tests would include observing
changes in the structure formation within such bubble \cite{Cooray:2002yx}. In fact, using
NVSS radio sources, it has been claimed that there is an anomalous
underdensity of radio sources coincidental with the position of the
cold spot in the CMB \cite{Rudnick:2007kw}, but this claim has been
discarded upon closer inspection \cite{Smith:2008tc}. However, Figure
\ref{fig:psidelta} shows that our scenario predicts a compensated bubble
wall, rather than a void, which provides a new (and unique) template
for comparison with galaxy surveys.

\section*{Acknowledgements}
We would like to thank Andrei Frolov, Ghazal Geshnizjani, and Tanmay Vachaspati  for useful discussions.
The authors acknowledge the hospitality of the Kavli Institute for Theoretical Physics in Beijing (KITPC), where this work was originated. The research at KITPC was supported in part by the Project of Knowledge Innovation Program (PKIP) of Chinese Academy of Sciences, Grant No. KJCX2.YW.W10.
NA is in part supported by Perimeter Institute (PI) for Theoretical Physics.  Research at PI is
supported by the Government of Canada through Industry Canada and by the
Province of Ontario through the Ministry of Research \& Innovation.
AS is supported in part by the U.S. Department of Energy under
Contract No. DE-AC02-98CH10886. YW is supported in part by Institute of Particle Physics, and by funds from McGill University.

\bibliographystyle{JHEP}
\bibliography{cosmo,cosmo_preprints,coldspot,void_nia}

\end{document}